\documentclass{article}
\usepackage{amssymb}
\usepackage{amsfonts}
\usepackage{amsmath}

\input{tcilatex}
\begin{document}

\title{On the asymptotic behavior of solutions to a class of grand canonical
master equations }
\author{Sabine B\"{o}gli$^{\ast }$ and Pierre-A. Vuillermot$^{\ast \ast
,\ast \ast \ast }$ \\
Department of Mathematical Sciences, Durham University,\\
Durham DH1 3LE, United Kingdom$^{\ast }$\\
sabine.boegli@durham.ac.uk\\
Universit\'{e} de Lorraine, CNRS, IECL, F-54000 Nancy, France$^{\ast \ast }$%
\\
Grupo de F\'{\i}sica Matem\'{a}tica, GFMUL, Faculdade de Ci\^{e}ncias,\\
Universidade de Lisboa, 1749-016 Lisboa, Portugal$^{\ast \ast \ast }$\\
pierre.vuillermot@univ-lorraine.fr}
\date{}
\maketitle

\begin{abstract}
In this article we investigate the long-time behavior of solutions to a
class of infinitely many master equations defined from transition rates that
are suitable for the description of a quantum system approaching
thermodynamical equilibrium with a heat bath at fixed temperature and a
reservoir consisting of one species of particles characterized by a fixed
chemical potential. We do so by proving a result which pertains to the
spectral resolution of the semigroup generated by the equations, whose
infinitesimal generator is realized as a trace-class self-adjoint operator
defined in a suitably weighted sequence space. This allows us to prove the
existence of global solutions which all stabilize toward the grand canonical
equilibrium probability distribution as the time variable becomes large,
some of them doing so exponentially rapidly. When we set the chemical
potential equal to zero, the stability statements continue to hold in the
sense that all solutions converge toward the Gibbs probability distribution
of the canonical ensemble which characterizes the equilibrium of the given
system with a heat bath at fixed temperature.

\ \ \ \ \ \ \ \ \ \ \ \ \ \ \ \ \ \ \ \ \ \ \ \ \ \ \ \ \ \ \ \ \ \ \ \ \ \ 
\textbf{Keywords: }

\ \ \ \ \ \ \ \ \ \ \ \ \ \ \ \ \textbf{Long Time Behavior, Master Equations}

\ \ \ \ \ \ \ \ \ \ \ \ \ \ \ \ \textbf{MSC 2020: 47.A.75 \ \ 47.B.93 \ \
47.D.06}

\ \ \ \ \ \ \ \ \ \ \ \ \ \ \ \ \ \ \ \ \ \ \ \ \ \ \ \ \ \ \ \ \ \ \textbf{%
Abbreviated Title:}

\ \ \ \ \ \ \ \ \ \ \ \ \ \ \textbf{\ \ \ Time evolution and master equations%
}

\textbf{\ \ \ \ \ \ \ \ \ \ \ \ \ \ \ \ \ \ \ \ \ \ \ }

\ \ \ \ \ \ \ \ \ \ \ \ \ \ \ \ \ \ \ \ \ 

\ \ \ \ \ \ \ \ \ \ \ \ \ \ \ \ \ 
\end{abstract}

\bigskip

\section{Introduction and outline}

It is well known that the grand canonical ensemble of statistical mechanics
provides a formalism suitable for the description of the properties of
classical or quantum systems in thermodynamical equilibrium with a heat bath
a fixed temperature and a reservoir of possibly different species of
particles, each of which being characterized by a chemical potential (see,
e.g., \cite{becker} and \cite{huang} for definitions and applications of the
above notions in various concrete situations). From the microscopic
properties of the systems it is then possible in principle to derive all
their macroscopic thermodynamical properties by means of the so-called grand
canonical partition function, which depends on the temperature and on the
chemical potentials we alluded to above. In order to achieve that for
systems that are not in thermodynamical equilibrium initially, an important
link may be provided by the solutions to certain master equations. In the
simplest setting of a system described by a Hamiltonian having a discrete
point spectrum, those solutions represent time-dependent probabilities which
determine the chance for jumps to occur between the various quantum states.
They also play a significant role in the stochastic approach to equilibrium
and non equilibrium thermodynamics of chemical reactions (see, e.g., the
theory and the applications developed in \cite{tomeoliveira}, \cite%
{tomeoliveirabis} and their numerous references, as well as in Chapter V of 
\cite{vankampen}. For the investigation of master equations in a different
or more general context with many important applications we also refer the
reader to \cite{breuerpetru}-\cite{glauber}, \cite{haake}, \cite%
{mozgunovlidar} and \cite{schnakenberg}).

It is precisely the long-time behavior of solutions to a class of various
initial-value problems for infinitely many master equations which is the
main theme of this article. The class in question is associated with
sequences of real numbers $\left( \lambda _{\mathsf{m}}\right) $ and of
non-negative integers $\left( \mathsf{N}_{\mathsf{m}}\right) $ indexed by $%
\mathsf{m}\in \mathbb{N}^{+}$, where the former may be interpreted for
instance as the point spectrum of some Hamiltonian and the latter as the
sequence of number of particles of a single species in the corresponding
quantum states. More specifically, we organize the remaining part of this
article in the following way: in Section 2 we define the relevant initial
value problems in which the transition rates depending on $\left( \lambda _{%
\mathsf{m}}\right) $ and $\left( \mathsf{N}_{\mathsf{m}}\right) $ are chosen
in such a way that the so-called detailed balance conditions of statistical
mechanics hold with respect to the grand canonical equilibrium probability
distribution. We then interpret the master equations as a dynamical system
defined on a suitable infinite-dimensional weighted sequence space, which
allows us to realize the infinitesimal generator of the system as a
trace-class self-adjoint operator whose spectral properties we investigate
in detail. In particular, we prove there a localization principle for all of
its eigenvalues and note the absence of a spectral gap around the zero
eigenvalue. This eventually leads us to the spectral resolution of the
corresponding semigroup whose consequences we analyze in Section 3, where we
show that the system of master equations we consider possesses global
solutions which all stabilize toward the grand canonical equilibrium
probability distribution as the time variable becomes large, some of them
doing so exponentially rapidly. In the important particular case where the
chemical potential is set equal to zero, the stability statements remain
true in that all solutions converge toward the Gibbs equilibrium probability
distribution of the canonical ensemble, some of them again exponentially
rapidly. Finally, we also consider there a concrete example involving the
quantum harmonic oscillator which shows how the decay properties of the
Fourier coefficients of the initial conditions can impact on the speed of
convergence of the solutions, ending up with power-law and even logarithmic
rates of decay.

We conclude this introduction by noting that the mere idea of making the
generator of a system of master equations a formally self-adjoint operator
by using the detailed balance conditions already appears as a set of remarks
scattered in Chapter V of \cite{vankampen}. As we shall see below, the
method of investigation we use in this article represents a systematic and
rigorous implementation of those remarks in a very specific context.

\section{On the spectral resolution of the semigroup generated by a class of
master equations}

As outlined in the introduction, we start out with a sequence of real
numbers $\left( \lambda _{\mathsf{m}}\right) $ and of non-negative integers $%
\left( \mathsf{N}_{\mathsf{m}}\right) $ indexed by $\mathsf{m}\in \mathbb{N}%
^{+}$, such that the grand canonical partition function satisfies%
\begin{equation}
\Theta _{\beta ,\mu }:=\dsum\limits_{\mathsf{m}=1}^{+\infty }\exp \left[
-\beta \left( \lambda _{\mathsf{m}}-\mu \mathsf{N}_{\mathsf{m}}\right) %
\right] <+\infty  \label{grandpartition}
\end{equation}%
for each $\beta >0$ and every $\mu \in \mathbb{R}$, where $\beta $ may be
interpreted as the inverse temperature and $\mu $ as the chemical potential.
By means of (\ref{grandpartition}) we then define the grand canonical
equilibrium probabilities by%
\begin{equation}
p_{\beta ,\mu ,\mathsf{m}}:=\Theta _{\beta ,\mu }^{-1}\exp \left[ -\beta
\left( \lambda _{\mathsf{m}}-\mu \mathsf{N}_{\mathsf{m}}\right) \right]
\label{equiprobabilities}
\end{equation}%
for each $\mathsf{m}\in \mathbb{N}^{+}$, and with every such $\mathsf{m}$ we
associate the class of initial-value problems for master equations of the
form%
\begin{eqnarray}
\frac{dp_{\mathsf{m}}\left( \tau \right) }{d\tau } &=&\sum_{\mathsf{n=1}%
}^{+\infty }\left( r_{\mathsf{m,n}}p_{\mathsf{n}}\left( \tau \right) -r_{%
\mathsf{n,m}}p_{\mathsf{m}}\left( \tau \right) \right) ,\text{ \ }\tau \in %
\left[ 0,+\infty \right) ,  \notag \\
p_{\mathsf{m}}\left( 0\right) &=&p_{\mathsf{m}}^{\ast }
\label{masterequation}
\end{eqnarray}%
where $\left( p_{\mathsf{m}}^{\ast }\right) $ stands for any sequence of
initial-data satisfying%
\begin{equation}
p_{\mathsf{m}}^{\ast }\geq 0\text{, \ }\sum_{\mathsf{m=1}}^{+\infty }\text{\ 
}p_{\mathsf{m}}^{\ast }=1.  \label{probabilities}
\end{equation}%
In (\ref{masterequation}) the transition rates $r_{\mathsf{m,n}}>0$ from
level $\mathsf{n}$ to level $\mathsf{m}$ are chosen in such a way that the
so-called detailed balance conditions%
\begin{equation}
r_{\mathsf{m,n}}p_{\beta ,\mu ,\mathsf{n}}=r_{\mathsf{n,m}}p_{\beta ,\mu ,%
\mathsf{m}}  \label{detailedbalance}
\end{equation}%
are satisfied for each $\beta >0$, every $\mu \in \mathbb{R}$ and all $%
\mathsf{m,n}\in \mathbb{N}^{+}$. In this manner the $p_{\beta ,\mu ,\mathsf{m%
}}$ provide a time-independent solution to (\ref{masterequation}) when we
choose $p_{\mathsf{m}}^{\ast }=p_{\beta ,\mu ,\mathsf{m}}$ for every $%
\mathsf{m}$, in addition to the fact that they make the corresponding
entropy production as defined in \cite{schnakenberg} equal to zero (see
also, e.g., Section II A in \cite{tomeoliveira} for a thorough discussion of
this point). Therefore, they do provide genuine equilibrium probabilities
indeed. Furthermore, owing to (\ref{equiprobabilities}) we may rewrite (\ref%
{detailedbalance}) as%
\begin{equation}
\frac{r_{\mathsf{m,n}}}{r_{\mathsf{n,m}}}=\exp \left[ -\beta \left( \lambda
_{\mathsf{m}}-\lambda _{\mathsf{n}}\right) +\beta \mu \left( \mathsf{N}_{%
\mathsf{m}}-\mathsf{N}_{\mathsf{n}}\right) \right] ,
\label{detailedbalancebis}
\end{equation}%
which is the starting point for the analysis of chemical reactions by means
of stochastic thermodynamics put forward in \cite{tomeoliveirabis} (see, in
particular, Section III of that article). In particular we may take%
\begin{equation}
r_{\mathsf{m,n}}=c_{\mathsf{m,n}}\exp \left[ -\frac{\beta }{2}\left( \lambda
_{\mathsf{m}}-\lambda _{\mathsf{n}}\right) +\frac{\beta \mu }{2}\left( 
\mathsf{N}_{\mathsf{m}}-\mathsf{N}_{\mathsf{n}}\right) \right]  \label{rates}
\end{equation}%
where the prefactors stand for any choice of real coefficients satisfying
the symmetry condition $c_{\mathsf{m,n}}=c_{\mathsf{n,m}}$ for all $\mathsf{%
m,n}\in \mathbb{N}^{+}$. In what follows we investigate (\ref{masterequation}%
) as a dynamical system on a suitable weighted sequence space with rates of
the form (\ref{rates}), which requires a specific and of course non unique
choice of the $c_{\mathsf{m,n}}$ to ensure that the dynamical system in
question be well defined. In fact, in order to keep our upcoming
computations as simple as possible we shall settle for%
\begin{equation}
c_{\mathsf{m,n}}=\exp \left[ -\beta \left( \lambda _{\mathsf{m}}+\lambda _{%
\mathsf{n}}\right) -\beta \mu \left( \mathsf{N}_{\mathsf{m}}+\mathsf{N}_{%
\mathsf{n}}\right) \right] ,  \label{suitableconstants}
\end{equation}%
which will play the role of convergence factors in Proposition 1 below as we
shall soon explain. Thus, let us denote by $l_{\mathbb{C},\mathsf{w}_{\beta
,\mu }}^{2}$the set of all complex sequences $\mathsf{p}:=$ $\left( p_{%
\mathsf{m}}\right) $ satisfying%
\begin{equation}
\left\Vert \mathsf{p}\right\Vert _{2,\mathsf{w}_{\beta ,\mu }}^{2}:=\sum_{%
\mathsf{m=1}}^{\mathsf{+\infty }}w_{\beta ,\mu ,\mathsf{m}}\left\vert p_{%
\mathsf{m}}\right\vert ^{2}<+\infty  \label{norm}
\end{equation}%
where $w_{\beta ,\mu ,\mathsf{m}}:=\exp \left[ \beta \left( \lambda _{%
\mathsf{m}}-\mu \mathsf{N}_{\mathsf{m}}\right) \right] $, which becomes a
complex separable Hilbert space when endowed with the usual operations and
the sesquilinear form%
\begin{equation}
\left( \mathsf{p},\mathsf{q}\right) _{2,\mathsf{w}_{\beta ,\mu }}:=\sum_{%
\mathsf{m=1}}^{\mathsf{+\infty }}w_{\beta ,\mu ,\mathsf{m}}p_{\mathsf{m}}%
\bar{q}_{\mathsf{m}}  \label{innerproduct}
\end{equation}%
defined with respect to the weight sequence $\mathsf{w}_{\beta ,\mu }\mathsf{%
:=}$ $\left( w_{\beta ,\mu ,\mathsf{m}}\right) $. Furthermore, let us
reformulate (\ref{masterequation}) as 
\begin{eqnarray}
\frac{dp_{\mathsf{m}}\left( \tau \right) }{d\tau } &=&\sum_{\mathsf{n=1}%
}^{+\infty }a_{\mathsf{m,n}}p_{\mathsf{n}}\left( \tau \right) ,\text{ \ }%
\tau \in \left[ 0,+\infty \right) ,  \notag \\
p_{\mathsf{m}}\left( 0\right) &=&p_{\mathsf{m}}^{\ast }\text{ \label%
{masterequationter}}
\end{eqnarray}%
where \ 
\begin{equation}
a_{\mathsf{m,n}}=\left\{ 
\begin{array}{c}
-\sum_{\mathsf{k=1},\text{ }\mathsf{k\neq m}}^{+\infty }r_{\mathsf{k,m}}%
\text{ \ \ for }\mathsf{m=n,} \\ 
\\ 
r_{\mathsf{m,n}}\text{\textsf{\ \ \ for }}\mathsf{m\neq n.}%
\end{array}%
\right.  \label{matrix}
\end{equation}

Then the following preliminary result holds, which is interesting in its own
right:

\bigskip

\textbf{Proposition 1. }\textit{For each} $\mathsf{p}\in $ $l_{\mathbb{C},%
\mathsf{w}_{\beta ,\mu }}^{2}$, \textit{the expression}%
\begin{equation}
\left( A\mathsf{p}\right) _{\mathsf{m}}:=\dsum\limits_{\mathsf{n}%
=1}^{+\infty }a_{\mathsf{m,n}}p_{\mathsf{n}}  \label{operator}
\end{equation}%
\textit{defines a linear, self-adjoint trace-class operator} $A:l_{\mathbb{C}%
,\mathsf{w}_{\beta ,\mu }}^{2}\mapsto l_{\mathbb{C},\mathsf{w}_{\beta ,\mu
}}^{2}$ \textit{whose trace is given by}%
\begin{equation}
\func{Tr}A=\Theta _{2\beta ,-\mu }-\Theta _{\frac{\beta }{2},-3\mu }\Theta _{%
\frac{3\beta }{2},-\frac{\mu }{3}}<0.  \label{trace}
\end{equation}%
\textbf{Proof. }We begin by showing that $A$ is a bounded operator.
Rewriting (\ref{operator}) as 
\begin{equation*}
\left( A\mathsf{p}\right) _{\mathsf{m}}=\dsum\limits_{\mathsf{n}=1}^{+\infty
}\left( a_{\mathsf{m,n}}w_{\beta ,\mu ,\mathsf{n}}^{-\frac{1}{2}}\right)
\left( w_{\beta ,\mu ,\mathsf{n}}^{\frac{1}{2}}p_{\mathsf{n}}\right)
\end{equation*}%
and using the Cauchy-Schwarz inequality we first obtain%
\begin{equation}
\left\Vert A\mathsf{p}\right\Vert _{2,\mathsf{w}_{\beta ,\mu }}^{2}\leqslant
\dsum\limits_{\mathsf{m}=1}^{+\infty }w_{\beta ,\mu ,\mathsf{m}%
}\dsum\limits_{\mathsf{n}=1}^{+\infty }w_{\beta ,\mu ,\mathsf{n}%
}^{-1}\left\vert a_{\mathsf{m,n}}\right\vert ^{2}\times \left\Vert \mathsf{p}%
\right\Vert _{2,\mathsf{w}_{\beta ,\mu }}^{2}.  \label{boundedoperator}
\end{equation}%
Furthermore, using (\ref{matrix}) we may write and estimate the right-hand
side in (\ref{boundedoperator}) as%
\begin{eqnarray}
&&\dsum\limits_{\mathsf{m}=1}^{+\infty }w_{\beta ,\mu ,\mathsf{m}%
}\dsum\limits_{\mathsf{n}=1}^{+\infty }w_{\beta ,\mu ,\mathsf{n}%
}^{-1}\left\vert a_{\mathsf{m,n}}\right\vert ^{2}  \notag \\
&=&\dsum\limits_{\mathsf{m}=1}^{+\infty }\left( \left\vert a_{\mathsf{m,m}%
}\right\vert ^{2}+w_{\beta ,\mu ,\mathsf{m}}\dsum\limits_{\mathsf{n}=1,%
\mathsf{n\neq m}}^{+\infty }w_{\beta ,\mu ,\mathsf{n}}^{-1}r_{\mathsf{m,n}%
}^{2}\right)  \label{estimates} \\
&\leqslant &\dsum\limits_{\mathsf{m}=1}^{+\infty }\left( \left( \sum_{%
\mathsf{n=1}}^{+\infty }r_{\mathsf{n,m}}\right) ^{2}+w_{\beta ,\mu ,\mathsf{m%
}}\dsum\limits_{\mathsf{n}=1}^{+\infty }w_{\beta ,\mu ,\mathsf{n}}^{-1}r_{%
\mathsf{m,n}}^{2}\right) .  \notag
\end{eqnarray}%
In addition, putting (\ref{suitableconstants}) into (\ref{rates}) gives%
\begin{equation}
r_{\mathsf{m,n}}=\exp \left[ -\frac{\beta }{2}\left( 3\lambda _{\mathsf{m}%
}+\mu \mathsf{N}_{\mathsf{m}}\right) -\frac{\beta }{2}\left( \lambda _{%
\mathsf{n}}+3\mu \mathsf{N}_{\mathsf{n}}\right) \right] ,  \label{ratesbis}
\end{equation}%
so that by taking (\ref{grandpartition}) and the expression for $w_{\beta
,\mu ,\mathsf{m}}$ into account we obtain%
\begin{equation*}
\sum_{\mathsf{n=1}}^{+\infty }r_{\mathsf{n,m}}=\Theta _{\frac{3\beta }{2},-%
\frac{\mu }{3}}\exp \left[ -\frac{\beta }{2}\left( \lambda _{\mathsf{m}%
}+3\mu \mathsf{N}_{\mathsf{m}}\right) \right]
\end{equation*}%
and%
\begin{equation*}
\dsum\limits_{\mathsf{n}=1}^{+\infty }w_{\beta ,\mu ,\mathsf{n}}^{-1}r_{%
\mathsf{m,n}}^{2}=\Theta _{2\beta ,-\mu }\exp \left[ -\beta \left( 3\lambda
_{\mathsf{m}}+\mu \mathsf{N}_{\mathsf{m}}\right) \right] .
\end{equation*}%
The substitution of these expressions into the last line of (\ref{estimates}%
) and a straightforward computation then lead to the estimate%
\begin{equation}
\dsum\limits_{\mathsf{m}=1}^{+\infty }w_{\beta ,\mu ,\mathsf{m}%
}\dsum\limits_{\mathsf{n}=1}^{+\infty }w_{\beta ,\mu ,\mathsf{n}%
}^{-1}\left\vert a_{\mathsf{m,n}}\right\vert ^{2}\leq \Theta _{\beta ,-3\mu
}\Theta _{\frac{3\beta }{2},-\frac{\mu }{3}}^{2}+\Theta _{2\beta ,-\mu
}^{2}<+\infty ,  \label{convergence}
\end{equation}%
which proves that $A$ is indeed a bounded operator.

Next, we observe that the detailed balance conditions (\ref{detailedbalance}%
) may be rewritten as%
\begin{equation*}
a_{\mathsf{m,n}}w_{\beta ,\mu ,\mathsf{m}}=a_{\mathsf{n,m}}w_{\beta ,\mu ,%
\mathsf{n}}
\end{equation*}%
for all $\mathsf{m,n}\in \mathbb{N}^{+}$, which immediately implies the
relation%
\begin{equation*}
\left( A\mathsf{p},\mathsf{q}\right) _{2,\mathsf{w}_{\beta ,\mu }}=\sum_{%
\mathsf{m,n=1}}^{\mathsf{+\infty }}w_{\beta ,\mu ,\mathsf{m}}a_{\mathsf{m,n}%
}p_{\mathsf{n}}\bar{q}_{\mathsf{m}}=\sum_{\mathsf{m,n=1}}^{\mathsf{+\infty }%
}w_{\beta ,\mu ,\mathsf{n}}a_{\mathsf{n,m}}p_{\mathsf{n}}\bar{q}_{\mathsf{m}%
}=\left( \mathsf{p},A\mathsf{q}\right) _{2,\mathsf{w}_{\beta ,\mu }}
\end{equation*}%
so that $A$ is self-adjoint.

In order to prove that $A$ is trace-class let us introduce the sequence of
canonical vectors $\left( \mathsf{e}_{\mathsf{m}}\right) $ given by $\left( 
\mathsf{e}_{\mathsf{m}}\right) _{\mathsf{n}}=\delta _{\mathsf{m,n}}$ for all 
$\mathsf{m,n}\in \mathbb{N}^{+},$ and let us consider the sequence defined
by $\mathsf{f}_{\mathsf{m}}$ $=w_{\beta ,\mu ,\mathsf{m}}^{-\frac{1}{2}}%
\mathsf{e}_{\mathsf{m}}$ for each $\mathsf{m}\in \mathbb{N}^{+}$. From this
and (\ref{innerproduct}) it follows immediately that the $\mathsf{f}_{%
\mathsf{m}}$ form an orthonormal system in $l_{\mathbb{C},\mathsf{w}_{\beta
,\mu }}^{2}$. Moreover we have $\left( \mathsf{f}_{\mathsf{m}},\mathsf{q}%
\right) _{2,\mathsf{w}_{,\beta ,\mu }}=w_{\beta ,\mu ,\mathsf{m}}^{\frac{1}{2%
}}\bar{q}_{\mathsf{m}}$ for every $\mathsf{q}$ $\in l_{\mathbb{C},\mathsf{w}%
_{\beta ,\mu }}^{2}$, so that if $\left( \mathsf{f}_{\mathsf{m}},\mathsf{q}%
\right) _{2,\mathsf{w}_{\beta ,\mu }}=0$ for each $\mathsf{m}$ then $\mathsf{%
q=0}$. Therefore the $\mathsf{f}_{\mathsf{m}}$ constitute an orthonormal
basis in $l_{\mathbb{C},\mathsf{w}_{\beta ,\mu }}^{2}$, and furthermore a
direct computation shows that%
\begin{equation}
\left( A\mathsf{f}_{\mathsf{m}},\mathsf{f}_{\mathsf{n}}\right) _{2,\mathsf{w}%
_{\beta ,\mu }}=w_{\beta ,\mu ,\mathsf{n}}^{\frac{1}{2}}a_{\mathsf{n,m}%
}w_{\beta ,\mu ,\mathsf{m}}^{-\frac{1}{2}}  \label{matrixelements}
\end{equation}%
for all $\mathsf{m,n}\in \mathbb{N}^{+}$. For any orthonormal basis $\left( 
\mathsf{g}_{\mathsf{m}}\right) $ in $l_{\mathbb{C},\mathsf{w}_{\beta ,\mu
}}^{2}$ we now have%
\begin{equation*}
A\mathsf{g}_{\mathsf{m}}=\dsum\limits_{\mathsf{j=1}}^{+\infty }\left( 
\mathsf{g}_{\mathsf{m}},\mathsf{f}_{\mathsf{j}}\right) _{2,\mathsf{w}_{\beta
,\mu }}A\mathsf{f}_{\mathsf{j}}
\end{equation*}%
after expanding each $\mathsf{g}_{\mathsf{m}}$ along the basis $\left( 
\mathsf{f}_{\mathsf{j}}\right) $. In this manner we obtain 
\begin{equation*}
\left( A\mathsf{g}_{\mathsf{m}},\mathsf{g}_{\mathsf{m}}\right) _{2,\mathsf{w}%
_{\beta ,\mu }}=\dsum\limits_{\mathsf{j,k=1}}^{+\infty }w_{\beta ,\mu ,%
\mathsf{k}}^{\frac{1}{2}}a_{\mathsf{k,j}}w_{\beta ,\mu ,\mathsf{j}}^{-\frac{1%
}{2}}\left( \mathsf{g}_{\mathsf{m}},\mathsf{f}_{\mathsf{j}}\right) _{2,%
\mathsf{w}_{\beta ,\mu }}\left( \mathsf{f}_{\mathsf{k}},\mathsf{g}_{\mathsf{m%
}}\right) _{2,\mathsf{w}_{\beta ,\mu }}
\end{equation*}%
according to (\ref{matrixelements}), so that the estimate%
\begin{eqnarray}
&&\dsum\limits_{\mathsf{m=1}}^{+\infty }\left\vert \left( A\mathsf{g}_{%
\mathsf{m}},\mathsf{g}_{\mathsf{m}}\right) _{2,\mathsf{w}_{\beta ,\mu
}}\right\vert  \notag \\
&\leqslant &\frac{1}{2}\dsum\limits_{\mathsf{j,k=1}}^{+\infty }w_{\beta ,\mu
,\mathsf{k}}^{\frac{1}{2}}\left\vert a_{\mathsf{k,j}}\right\vert w_{\beta
,\mu ,\mathsf{j}}^{-\frac{1}{2}}\dsum\limits_{\mathsf{m=1}}^{+\infty }\left(
\left\vert \left( \mathsf{g}_{\mathsf{m}},\mathsf{f}_{\mathsf{j}}\right) _{2,%
\mathsf{w}_{\beta ,\mu }}\right\vert ^{2}+\left\vert \left( \mathsf{g}_{%
\mathsf{m}},\mathsf{f}_{\mathsf{k}}\right) _{2,\mathsf{w}_{\beta ,\mu
}}\right\vert ^{2}\right)  \label{estimatesbis} \\
&=&\dsum\limits_{\mathsf{j,k=1}}^{+\infty }w_{\beta ,\mu ,\mathsf{k}}^{\frac{%
1}{2}}\left\vert a_{\mathsf{k,j}}\right\vert w_{\beta ,\mu ,\mathsf{j}}^{-%
\frac{1}{2}}  \notag
\end{eqnarray}%
holds. The last equality in (\ref{estimatesbis}) follows from the expansion
of each $\mathsf{f}_{\mathsf{j}}$ along the basis $\left( \mathsf{g}_{%
\mathsf{m}}\right) $, which entails the relation%
\begin{equation*}
\dsum\limits_{\mathsf{m=1}}^{+\infty }\left\vert \left( \mathsf{g}_{\mathsf{m%
}},\mathsf{f}_{\mathsf{j}}\right) _{2,\mathsf{w}_{\beta ,\mu }}\right\vert
^{2}=\left\Vert \mathsf{f}_{\mathsf{j}}\right\Vert _{2,\mathsf{w}_{\beta
,\mu }}^{2}=1
\end{equation*}%
for every $\mathsf{j}\in \mathbb{N}^{+}$. According to (\ref{matrix}) we
then have%
\begin{eqnarray*}
&&\dsum\limits_{\mathsf{k=1}}^{+\infty }w_{\beta ,\mu ,\mathsf{k}}^{\frac{1}{%
2}}\dsum\limits_{\mathsf{j=1}}^{+\infty }\left\vert a_{\mathsf{k,j}%
}\right\vert w_{\beta ,\mu ,\mathsf{j}}^{-\frac{1}{2}} \\
&=&\dsum\limits_{\mathsf{k}=1}^{+\infty }\left( \left\vert a_{\mathsf{k,k}%
}\right\vert +w_{\beta ,\mu ,\mathsf{k}}^{\frac{1}{2}}\dsum\limits_{\mathsf{%
j=1,j\neq k}}^{+\infty }r_{\mathsf{k,j}}w_{\beta ,\mu ,\mathsf{j}}^{-\frac{1%
}{2}}\right) \\
&\leqslant &\dsum\limits_{\mathsf{k}=1}^{+\infty }\left( \sum_{\mathsf{j=1}%
}^{+\infty }r_{\mathsf{j,k}}+w_{\beta ,\mu ,\mathsf{k}}^{\frac{1}{2}%
}\dsum\limits_{\mathsf{j}=1}^{+\infty }r_{\mathsf{k,j}}w_{\beta ,\mu ,%
\mathsf{j}}^{-\frac{1}{2}}\right) <+\infty
\end{eqnarray*}%
for the right-hand side of the equality in (\ref{estimatesbis}), where we
used (\ref{ratesbis}) and computations similar to those leading to (\ref%
{convergence}) to prove convergence. The series%
\begin{equation*}
\dsum\limits_{\mathsf{m=1}}^{+\infty }\left( A\mathsf{g}_{\mathsf{m}},%
\mathsf{g}_{\mathsf{m}}\right) _{2,\mathsf{w}_{\beta ,\mu }}
\end{equation*}%
is therefore itself convergent and since the orthonormal basis $\left( 
\mathsf{g}_{\mathsf{m}}\right) $ was arbitrary we may conclude that $A$ is
trace-class, with%
\begin{equation*}
\func{Tr}A=\dsum\limits_{\mathsf{m=1}}^{+\infty }\left( A\mathsf{f}_{\mathsf{%
m}},\mathsf{f}_{\mathsf{m}}\right) _{2,\mathsf{w}_{\beta ,\mu
}}=\dsum\limits_{\mathsf{m=1}}^{+\infty }a_{\mathsf{m,m}}=-\dsum\limits_{%
\mathsf{m=1}}^{+\infty }\sum_{\mathsf{n=1},\text{ }\mathsf{n\neq m}%
}^{+\infty }r_{\mathsf{n,m}}
\end{equation*}%
as a consequence of (\ref{matrix}) and (\ref{matrixelements}), which
eventually leads to (\ref{trace}). \ \ $\blacksquare $

\bigskip

\textsc{Remark.} Had we chosen (\ref{rates}) for the rates with $c_{\mathsf{%
m,n}}=1$ for all $\mathsf{m,n}\in \mathbb{N}^{+}$ instead of (\ref{ratesbis}%
), some of the series in the proof of Proposition 1 would have been
divergent, for instance the very last series on the right-hand side of (\ref%
{estimates}). That is the reason why we referred to (\ref{suitableconstants}%
) as convergence factors.

\bigskip

In what follows we state and prove the main result of this section, in which
we investigate in detail the spectral properties of $A$ including in
particular a principle of localization of the eigenvalues, from which we
obtain the spectral resolution of the semigroup generated by $A$. In this
context the sequence $\left( b_{\mathsf{m}}\right) $ given by%
\begin{equation}
b_{\mathsf{m}}=\Theta _{\frac{3\beta }{2},-\frac{\mu }{3}}\exp \left[ -\frac{%
\beta }{2}\left( \lambda _{\mathsf{m}}+3\mu \mathsf{N}_{\mathsf{m}}\right) %
\right]  \label{sequence}
\end{equation}%
plays an important role.

\bigskip

\textbf{Theorem 1.} \textit{Let }$A$\textit{\ be the operator defined by (%
\ref{operator}). Then the spectrum of }$A$, $\sigma (A)$,\textit{\ is a
discrete compact set with infinitely many real elements }$\left( \nu _{%
\mathsf{k}}\right) $ \textit{indexed by} $\mathsf{k}\in \mathbb{N}^{+}$, 
\textit{which are all eigenvalues including }$\nu _{1}=0$\textit{. }

\textit{Assuming in addition that}%
\begin{equation}
\lambda _{\mathsf{m+1}}-\lambda _{\mathsf{m}}>3\mu \left( \mathsf{N}_{%
\mathsf{m}}-\mathsf{N}_{\mathsf{m+1}}\right)  \label{spectralcondition}
\end{equation}%
\textit{for every} $\mathsf{m}$\textsf{\ }$\in \mathbb{N}^{+}$,\textit{\ the
following two statements also hold:}

\textit{(a) Each eigenvalue of }$A$ \textit{is simple and the corresponding
eigenspace is spanned by }$\mathsf{\hat{p}}_{\mathsf{k}}=\left( \hat{p}_{%
\mathsf{k,m}}\right) $ \textit{where}%
\begin{equation*}
\hat{p}_{\mathsf{k,m}}=\frac{\exp \left[ -\frac{\beta }{2}\left( 3\lambda _{%
\mathsf{m}}+\mu \mathsf{N}_{\mathsf{m}}\right) \right] }{\nu _{\mathsf{k}%
}+b_{\mathsf{m}}}.
\end{equation*}%
\textit{Moreover, each such an eigenvalue} \textit{is implicitly
characterized by the relation}%
\begin{equation}
\dsum\limits_{\mathsf{m}=1}^{+\infty }\frac{\exp \left[ -\beta \left(
3\lambda _{\mathsf{m}}+\mu \mathsf{N}_{\mathsf{m}}\right) \right] }{\nu _{%
\mathsf{k}}+b_{\mathsf{m}}}=1.  \label{characterization}
\end{equation}%
\textit{Furthermore, the set of normalized eigenvectors given by}%
\begin{equation*}
\mathsf{\hat{q}}_{\mathsf{k}}:=\frac{\mathsf{\hat{p}}_{\mathsf{k}}}{%
\left\Vert \mathsf{\hat{p}}_{\mathsf{k}}\right\Vert _{2,\mathsf{w}_{\beta
,\mu }}}
\end{equation*}%
\textit{for every }$\mathsf{k}\in $ $\mathbb{N}^{+}$ \textit{constitutes an
orthonormal basis of }$l_{\mathbb{C},\mathsf{w}_{\beta ,\mu }}^{2}$\textit{.}

\textit{(b) If the nonzero eigenvalues of }$A$\textit{\ are ordered as }$\nu
_{\mathsf{k}}<\nu _{\mathsf{k+1}}$\textit{\ for every} $\mathsf{k}\in
\left\{ 2,3,...\right\} $, \textit{then they are localized according to}%
\begin{equation}
\nu _{\mathsf{k}}\in \left( -b_{\mathsf{k-1}},-b_{\mathsf{k}}\right)
\label{localization}
\end{equation}%
\textit{for every such }$k.$\textit{\ In particular, all the nonzero
elements of }$\sigma (A)$\textit{\ are negative and furthermore, for every }$%
\mathsf{p}\in $ $l_{\mathbb{C},\mathsf{w}_{\beta ,\mu }}^{2}$ \textit{we
have the norm-convergent spectral resolution}%
\begin{equation}
\exp \left[ \tau A\right] \mathsf{p=}\dsum\limits_{\mathsf{k=1}}^{+\infty
}\left( \mathsf{p,\hat{q}}_{\mathsf{k}}\right) _{2,\mathsf{w}_{\beta ,\mu
}}\exp \left[ \tau \nu _{\mathsf{k}}\right] \mathsf{\hat{q}}_{\mathsf{k}}
\label{spectralresolution}
\end{equation}%
\textit{of the semigroup }$\exp \left[ \tau A\right] _{\tau \in \left[
0,+\infty \right) \text{ }}$\textit{generated by }$A.$

\bigskip

\textbf{Proof.} From (\ref{equiprobabilities}) it is straightforward to
check that $\mathsf{p}_{\beta \mathsf{,}\mu }:=\left( p_{\beta ,\mu ,\mathsf{%
m}}\right) \in l_{\mathbb{C},\mathsf{w}_{\beta ,\mu }}^{2}$. Moreover, we
infer from Proposition 1 that $A$ is a compact self-adjoint operator in $l_{%
\mathbb{C},\mathsf{w}_{\beta ,\mu }}^{2}$, which implies in particular the
very first statement of the theorem since we have%
\begin{equation*}
A\mathsf{p}_{\beta \mathsf{,}\mu }=0
\end{equation*}%
as a consequence of (\ref{detailedbalance}), (\ref{matrix}) and (\ref%
{operator}).

As for the proof of Statement (a), we first note that the eigenvalue equation%
\begin{equation*}
A\mathsf{p}=\nu _{\mathsf{k}}\mathsf{p}
\end{equation*}%
is equivalent to having the relation%
\begin{equation}
\dsum\limits_{\mathsf{n=1}}^{+\infty }r_{\mathsf{m,n}}p_{\mathsf{n}}=\left(
\nu _{\mathsf{k}}+\dsum\limits_{\mathsf{n=1}}^{+\infty }r_{\mathsf{n,m}%
}\right) p_{\mathsf{m}}  \label{identity}
\end{equation}%
satisfied for all $\mathsf{m,k}\in \mathbb{N}^{+}$. We then use (\ref%
{ratesbis}) in (\ref{identity}) to get%
\begin{equation}
\mathsf{c}_{\mathsf{p},\beta ,\mu }\exp \left[ -\frac{\beta }{2}\left(
3\lambda _{\mathsf{m}}+\mu \mathsf{N}_{\mathsf{m}}\right) \right] =\left(
\nu _{\mathsf{k}}+b_{\mathsf{m}}\right) p_{\mathsf{m}}  \label{identitybis}
\end{equation}%
where we took (\ref{sequence}) into account and defined%
\begin{equation}
\mathsf{c}_{\mathsf{p},\beta ,\mu }:=\dsum\limits_{\mathsf{n=1}}^{+\infty
}\exp \left[ -\frac{\beta }{2}\left( 3\lambda _{\mathsf{n}}+\mu \mathsf{N}_{%
\mathsf{n}}\right) \right] p_{\mathsf{n}}.  \label{abreviation}
\end{equation}%
Now for any $\mathsf{p}\in l_{\mathbb{C},\mathsf{w}_{\beta ,\mu }}^{2}$ we
evidently have either $\mathsf{c}_{\mathsf{p},\beta ,\mu }\neq 0$ or $%
\mathsf{c}_{\mathsf{p},\beta ,\mu }=0$. In the first case Relation (\ref%
{identitybis}) implies that $\left( \nu _{\mathsf{k}}+b_{\mathsf{m}}\right)
p_{\mathsf{m}}\neq 0$ for each $\mathsf{m}\in \mathbb{N}^{+}$, so that we
may solve for $p_{\mathsf{m}}$ and get 
\begin{equation}
p_{\mathsf{k,m}}=\mathsf{c}_{\mathsf{p}_{\mathsf{k}},\beta ,\mu }\hat{p}_{%
\mathsf{k,m}}  \label{eigenvectors}
\end{equation}%
where%
\begin{equation}
\hat{p}_{\mathsf{k,m}}:=\frac{\exp \left[ -\frac{\beta }{2}\left( 3\lambda _{%
\mathsf{m}}+\mu \mathsf{N}_{\mathsf{m}}\right) \right] }{\nu _{\mathsf{k}%
}+b_{\mathsf{m}}}.  \label{eigenvectorsbis}
\end{equation}%
Moreover, with the $\hat{p}_{\mathsf{k,m}}$ given by (\ref{eigenvectorsbis})
we claim that $\mathsf{\hat{p}}_{\mathsf{k}}:\mathsf{=}\left( \hat{p}_{%
\mathsf{k,m}}\right) \in l_{\mathbb{C},\mathsf{w}_{\beta ,\mu }}^{2}$. On
the one hand, this is clear if $\nu _{\mathsf{k}}=0$ for then (\ref%
{eigenvectorsbis}) reduces to $\mathsf{p}_{\beta \mathsf{,}\mu }$ up to a
trivial multiplicative constant. On the other hand, if $\nu _{\mathsf{k}%
}\neq 0$ we have%
\begin{eqnarray}
&&\sum_{\mathsf{m=1}}^{\mathsf{+\infty }}w_{\beta ,\mu ,\mathsf{m}%
}\left\vert \nu _{\mathsf{k}}+b_{\mathsf{m}}\right\vert ^{2}\left\vert \hat{p%
}_{\mathsf{k,m}}\right\vert ^{2}  \notag \\
&=&\sum_{\mathsf{m=1}}^{\mathsf{+\infty }}\exp \left[ -2\beta \left( \lambda
_{\mathsf{m}}+\mu \mathsf{N}_{\mathsf{m}}\right) \right] =\Theta _{2\beta
,-\mu }<+\infty  \label{convergencebis}
\end{eqnarray}%
from (\ref{eigenvectorsbis}) and (\ref{grandpartition}), the latter also
implying that $\lim_{\mathsf{m}\rightarrow +\infty }b_{\mathsf{m}}=0$.
Therefore we have%
\begin{equation*}
\lim_{\mathsf{m}\rightarrow +\infty }\left\vert \nu _{\mathsf{k}}+b_{\mathsf{%
m}}\right\vert =\left\vert \nu _{\mathsf{k}}\right\vert \neq 0
\end{equation*}%
so that (\ref{convergencebis}) implies%
\begin{equation*}
\sum_{\mathsf{m=1}}^{\mathsf{+\infty }}w_{\beta ,\mu ,\mathsf{m}}\left\vert 
\hat{p}_{\mathsf{k,m}}\right\vert ^{2}<+\infty
\end{equation*}%
by asymptotic comparison, as desired. In this manner the $\mathsf{\hat{p}}_{%
\mathsf{k}}$ provide a set of eigenvectors of $A$ associated with the $\nu _{%
\mathsf{k}}$, and we now prove that there are no others. Indeed, in the
second case we alluded to above where $\mathsf{c}_{\mathsf{p},\beta ,\mu }=0$%
, we have $\left( \nu _{\mathsf{k}}+b_{\mathsf{m}}\right) p_{\mathsf{m}}=0$
for each $\mathsf{m}\in \mathbb{N}^{+}$ and therefore there exists an $%
\mathsf{m}^{\ast }\in \mathbb{N}^{+}$ such that $\nu _{\mathsf{k}}+b_{%
\mathsf{m}^{\ast }}=0$ since $\mathsf{p}=0$ is not an eigenvector. But the
spectral condition (\ref{spectralcondition}) is equivalent to having $b_{%
\mathsf{m}+1}<b_{\mathsf{m}}$ for each $\mathsf{m}\in \mathbb{N}^{+}$, so
that the $\mathsf{m}^{\ast }$ in question is unique. Consequently we
necessarily have $p_{\mathsf{m}}=0$ for every $\mathsf{m\neq m}^{\ast }$ and 
$p_{\mathsf{m}^{\ast }}\neq 0$, which implies the relation%
\begin{equation*}
\mathsf{c}_{\mathsf{p},\beta ,\mu }=\exp \left[ -\frac{\beta }{2}\left(
3\lambda _{\mathsf{m}^{\ast }}+\mu \mathsf{N}_{\mathsf{m}^{\ast }}\right) %
\right] p_{\mathsf{m}^{\ast }}\neq 0,
\end{equation*}%
a contradiction. Finally, the characterization (\ref{characterization}) of
the eigenvalues is a direct consequence of the substitution of (\ref%
{eigenvectors}) into (\ref{abreviation}). The preceding considerations thus
prove the first part of Statement (a), while the second part follows
immediately from the fact that $A$ is a compact self-adjoint operator.

Let us now prove Statement (b) by first ordering the non-zero eigenvalues of 
$A$ as $\nu _{\mathsf{k}}<\nu _{\mathsf{k+1}}$\ for every $\mathsf{k}\in
\left\{ 2,3,...\right\} $. To this end we consider the auxiliary function $%
\mathsf{a}$ $:\left( -\infty ,0\right) \setminus \left\{ -b_{\mathsf{m}},%
\text{ }\mathsf{m}\in \mathbb{N}^{+}\right\} $ defined by%
\begin{equation*}
\mathsf{a}(\nu ):=\dsum\limits_{\mathsf{m}=1}^{+\infty }\frac{\exp \left[
-\beta \left( 3\lambda _{\mathsf{m}}+\mu \mathsf{N}_{\mathsf{m}}\right) %
\right] }{\nu +b_{\mathsf{m}}},
\end{equation*}%
and remark that this series is absolutely convergent by virtue of (\ref%
{grandpartition}) and the fact that $b_{\mathsf{m}}\rightarrow 0$ as $%
\mathsf{m}\rightarrow +\infty $. Furthermore, it is easily verified that%
\begin{eqnarray*}
\lim_{\nu \searrow -b_{\mathsf{k-1}}}\mathsf{a}(\nu ) &=&+\infty , \\
\lim_{\nu \nearrow -b_{\mathsf{k}}}\mathsf{a}(\nu ) &=&-\infty ,
\end{eqnarray*}%
and that $\mathsf{a}^{\prime }(\nu )<0$ for every $\nu \in \left( -b_{%
\mathsf{k-1}},-b_{\mathsf{k}}\right) $, which implies the existence of a
unique $\nu _{\mathsf{k}}^{\ast }\in \left( -b_{\mathsf{k-1}},-b_{\mathsf{k}%
}\right) $ satisfying $\mathsf{a}(\nu _{\mathsf{k}}^{\ast })=1.$ Therefore,
from the characterization (\ref{characterization}) of the eigenvalues we
necessarily have $\nu _{\mathsf{k}}^{\ast }=\nu _{\mathsf{k}}$ for every $%
\mathsf{k}\in \left\{ 2,3,...\right\} $, thereby proving the first part of
Statement (b). Finally, for every $\mathsf{p}\in l_{\mathbb{C},\mathsf{w}%
_{\beta ,\mu }}^{2}$we have the norm-convergent expansion%
\begin{equation*}
\mathsf{p=}\dsum\limits_{\mathsf{k=1}}^{+\infty }\left( \mathsf{p,\hat{q}}_{%
\mathsf{k}}\right) _{2,\mathsf{w}_{\beta ,\mu }}\mathsf{\hat{q}}_{\mathsf{k}}
\end{equation*}%
from the last part of Statement (a), which implies (\ref{spectralresolution}%
) at once. \ \ $\blacksquare $

\bigskip

\textsc{Remark.} Since $A$ is trace-class, it follows from Lidski\u{\i}'s
theorem (see, e.g., Theorem 8.4 in Chapter III of \cite{gohbergkrein}) that
the so-called matrix trace (\ref{trace}) coincides with the spectral trace,
to wit,%
\begin{equation*}
\dsum\limits_{\mathsf{k=1}}^{+\infty }\nu _{\mathsf{k}}=\Theta _{2\beta
,-\mu }-\Theta _{\frac{\beta }{2},-3\mu }\Theta _{\frac{3\beta }{2},-\frac{%
\mu }{3}},
\end{equation*}%
which implies that $\lim_{\mathsf{k\rightarrow +\infty }}\nu _{\mathsf{k}%
}=0=\nu _{\mathsf{1}}$. Therefore, there is no spectral gap around the zero
eigenvalue of $A$ whose eigenspace is generated by $\mathsf{p}_{\beta 
\mathsf{,}\mu }$. In the next section we investigate some consequences of
this fact.

\section{On the eigenspace associated with the zero eigenvalue of $A$ as a
global attractor}

Since $\nu _{\mathsf{1}}=0$ is an accumulation point of $\sigma \left(
A\right) $, we might want to truncate expansion (\ref{spectralresolution})
in order to get an exponential decay of some sort for the solutions to (\ref%
{masterequation}), or else proceed more generally to obtain convergence
statements without error bounds, or more specifically with bounds that may
be slower than exponential. We first make the idea of truncation precise by
writing $\vee _{\mathsf{k=1}}^{\mathsf{N}}E_{\nu _{\mathsf{k}}}\left(
A\right) $ for the closed linear hull of $\cup _{\mathsf{k=1}}^{\mathsf{N}%
}E_{\nu _{\mathsf{k}}}\left( A\right) $ in $l_{\mathbb{C},\mathsf{w}_{\beta
,\mu }}^{2}$ for any $\mathsf{N}\in \mathbb{N}^{+}$, where $E_{\nu _{\mathsf{%
k}}}\left( A\right) $ stands for the eigenspace of $A$ associated with the
eigenvalue $\nu _{\mathsf{k}}$. Then we have:

\bigskip

\textbf{Theorem 2. }\textit{Let }$A$\textit{\ be the operator defined by (%
\ref{operator}), and let }$\mathsf{p}^{\ast }\in $ $l_{\mathbb{C},\mathsf{w}%
_{\beta ,\mu }}^{2}$ \textit{be any initial condition satisfying (\ref%
{probabilities}). Then }%
\begin{equation}
\left( \exp \left[ \tau A\right] \mathsf{p}^{\ast }\right) _{\mathsf{m}%
}\geqslant 0,\text{ \ }\sum_{\mathsf{m=1}}^{\mathsf{+\infty }}\left( \exp %
\left[ \tau A\right] \mathsf{p}^{\ast }\right) _{\mathsf{m}}\text{\ }=1
\label{probabilitiesbis}
\end{equation}%
\textit{for every }$\tau \in \left[ 0,+\infty \right) $.

\textit{Assuming moreover that (\ref{spectralcondition}) holds, and that the
ordering }$\nu _{\mathsf{k}}<\nu _{\mathsf{k+1}}$\textit{\ for every} $%
\mathsf{k}\in \left\{ 2,3,...\right\} $ \textit{is still valid,} \textit{%
then for each }$\mathsf{N}\in \mathbb{N}^{+}$ \textit{with} $\mathsf{%
N\geqslant 2}$ \textit{and} $\mathsf{p}^{\ast }\in \vee _{\mathsf{k=1}}^{%
\mathsf{N}}E_{\nu _{\mathsf{k}}}\left( A\right) $ \textit{satisfying (\ref%
{probabilities}) we have the exponential decay estimate}%
\begin{equation}
\left\Vert \exp \left[ \tau A\right] \mathsf{p}^{\ast }-\mathsf{p}_{\beta 
\mathsf{,}\mu }\right\Vert _{2,\mathsf{w}_{\beta ,\mu }}\mathsf{\leqslant }%
\exp \left[ -\tau \left\vert \nu _{\mathsf{N}}\right\vert \right] \left\Vert 
\mathsf{p}^{\ast }\right\Vert _{2,\mathsf{w}_{\beta ,\mu }}
\label{exponentialdecay}
\end{equation}%
\textit{for every }$\tau \in \left[ 0,+\infty \right) $, \textit{where} $%
\mathsf{p}_{\beta \mathsf{,}\mu }$ \textit{is given by (\ref%
{equiprobabilities})}.

\bigskip

\textbf{Proof.} Relations (\ref{probabilitiesbis}) are an immediate
consequence of some continuity arguments and of the summation of (\ref%
{masterequation}) over $\mathsf{m}\in \mathbb{N}^{+}$.

As for the proof of (\ref{exponentialdecay}), we start out from (\ref%
{spectralresolution}) to get%
\begin{equation*}
\exp \left[ \tau A\right] \mathsf{p}^{\ast }-\left( \mathsf{p}^{\ast }%
\mathsf{,\hat{q}}_{\mathsf{1}}\right) _{2,\mathsf{w}_{\beta ,\mu }}\mathsf{%
\hat{q}}_{\mathsf{1}}\mathsf{=}\dsum\limits_{\mathsf{k=2}}^{\mathsf{N}%
}\left( \mathsf{p}^{\ast }\mathsf{,\hat{q}}_{\mathsf{k}}\right) _{2,\mathsf{w%
}_{\beta ,\mu }}\exp \left[ \tau \nu _{\mathsf{k}}\right] \mathsf{\hat{q}}_{%
\mathsf{k}}
\end{equation*}%
since $\mathsf{p}^{\ast }$ is orthogonal to $\mathsf{\hat{q}}_{\mathsf{k}}$
in $l_{\mathbb{C},\mathsf{w}_{\beta ,\mu }}^{2}$ for each $\mathsf{%
k\geqslant N+1}$, so that from Parseval's relation we obtain%
\begin{equation}
\left\Vert \exp \left[ \tau A\right] \mathsf{p}^{\ast }-\left( \mathsf{p}%
^{\ast }\mathsf{,\hat{q}}_{\mathsf{1}}\right) _{2,\mathsf{w}_{\beta ,\mu }}%
\mathsf{\hat{q}}_{\mathsf{1}}\right\Vert _{2,\mathsf{w}_{\beta ,\mu }}^{2}%
\mathsf{\leqslant }\exp \left[ -2\tau \left\vert \nu _{\mathsf{N}%
}\right\vert \right] \left\Vert \mathsf{p}^{\ast }\right\Vert _{2,\mathsf{w}%
_{\beta ,\mu }}^{2}  \label{exponentialdecaybis}
\end{equation}%
for every $\tau \in \left[ 0,+\infty \right) $. It remains to show that%
\begin{equation}
\left( \mathsf{p}^{\ast }\mathsf{,\hat{q}}_{\mathsf{1}}\right) _{2,\mathsf{w}%
_{\beta ,\mu }}\mathsf{\hat{q}}_{\mathsf{1}}=\mathsf{p}_{\beta \mathsf{,}\mu
}.  \label{basicrelation}
\end{equation}%
From (\ref{equiprobabilities}) and (\ref{norm}) we first have%
\begin{equation*}
\left\Vert \mathsf{p}_{\beta \mathsf{,}\mu }\right\Vert _{2,\mathsf{w}%
_{\beta ,\mu }}^{2}=\Theta _{\beta ,\mu }^{-2}\dsum\limits_{\mathsf{m=1}}^{%
\mathsf{+\infty }}\exp \left[ -\beta \left( \lambda _{\mathsf{m}}-\mu 
\mathsf{N}_{\mathsf{m}}\right) \right] =\Theta _{\beta ,\mu }^{-1}
\end{equation*}%
as a consequence of (\ref{grandpartition}), so that we may choose $\mathsf{%
\hat{q}}_{\mathsf{1}}=\Theta _{\beta ,\mu }^{\frac{1}{2}}\mathsf{p}_{\beta 
\mathsf{,}\mu }$ as one of the unit eigenvectors associated with $\nu _{%
\mathsf{1}}=0$. Moreover, using (\ref{norm}) on the left-hand side of (\ref%
{exponentialdecaybis}) we eventually get%
\begin{equation*}
\left\vert \left( \exp \left[ \tau A\right] \mathsf{p}^{\ast }\right) _{%
\mathsf{m}}-\left( \mathsf{p}^{\ast }\mathsf{,\hat{q}}_{\mathsf{1}}\right)
_{2,\mathsf{w}_{\beta ,\mu }}\mathsf{\hat{q}}_{\mathsf{1,m}}\right\vert 
\mathsf{\leqslant }\exp \left[ -\frac{\beta }{2}\left( \lambda _{\mathsf{m}%
}-\mu \mathsf{N}_{\mathsf{m}}\right) \right] \exp \left[ -\tau \left\vert
\nu _{\mathsf{N}}\right\vert \right] \left\Vert \mathsf{p}^{\ast
}\right\Vert _{2,\mathsf{w}_{\beta ,\mu }}
\end{equation*}%
for every $\mathsf{m}$. Therefore, the summation of both sides of this
expression over $\mathsf{m}\in \mathbb{N}^{+}$ leads to%
\begin{equation*}
\left\vert 1-\left( \mathsf{p}^{\ast }\mathsf{,\hat{q}}_{\mathsf{1}}\right)
_{2,\mathsf{w}_{\beta ,\mu }}\dsum\limits_{\mathsf{m=1}}^{\mathsf{+\infty }}%
\mathsf{\hat{q}}_{\mathsf{1,m}}\right\vert \mathsf{\leqslant }\Theta _{\frac{%
\beta }{2},\mu }\exp \left[ -\tau \left\vert \nu _{\mathsf{N}}\right\vert %
\right] \left\Vert \mathsf{p}^{\ast }\right\Vert _{2,\mathsf{w}_{\beta ,\mu
}}
\end{equation*}%
where we have used (\ref{grandpartition}) and the normalization condition in
(\ref{probabilitiesbis}), so that letting $\tau \rightarrow +\infty $ in the
preceding relation necessarily gives%
\begin{equation}
\left( \mathsf{p}^{\ast }\mathsf{,\hat{q}}_{\mathsf{1}}\right) _{2,\mathsf{w}%
_{\beta ,\mu }}\dsum\limits_{\mathsf{m=1}}^{\mathsf{+\infty }}\mathsf{\hat{q}%
}_{\mathsf{1,m}}=1.  \label{normalization}
\end{equation}%
But from our choice of $\mathsf{\hat{q}}_{1}$ we have%
\begin{equation*}
\dsum\limits_{\mathsf{m=1}}^{\mathsf{+\infty }}\mathsf{\hat{q}}_{\mathsf{1,m}%
}=\Theta _{\beta ,\mu }^{\frac{1}{2}}\dsum\limits_{\mathsf{m=1}}^{\mathsf{%
+\infty }}\mathsf{p}_{\beta \mathsf{,}\mu ,\mathsf{m}}=\Theta _{\beta ,\mu
}^{\frac{1}{2}}
\end{equation*}%
and thereby%
\begin{equation*}
\left( \mathsf{p}^{\ast }\mathsf{,\hat{q}}_{\mathsf{1}}\right) _{2,\mathsf{w}%
_{\beta ,\mu }}=\Theta _{\beta ,\mu }^{-\frac{1}{2}}
\end{equation*}%
independently of $\mathsf{p}^{\ast }$. Consequently we end up with%
\begin{equation*}
\left( \mathsf{p}^{\ast }\mathsf{,\hat{q}}_{\mathsf{1}}\right) _{2,\mathsf{w}%
_{\beta ,\mu }}\mathsf{\hat{q}}_{\mathsf{1}}=\Theta _{\beta ,\mu }^{-\frac{1%
}{2}}\mathsf{\hat{q}}_{\mathsf{1}}=\mathsf{p}_{\beta \mathsf{,}\mu },
\end{equation*}%
as desired. \ \ $\blacksquare $

\bigskip

\textsc{Remark.} All the $\mathsf{p}^{\ast }\in \vee _{\mathsf{k=1}}^{%
\mathsf{N}}E_{\nu _{\mathsf{k}}}\left( A\right) $ satisfying (\ref%
{probabilities}) provide a large supply of initial data for which estimate (%
\ref{exponentialdecay}) holds, which obviously grows with $\mathsf{N}$. But
this is at the expense of having a smaller exponential rate of decay
whenever $\mathsf{N}$ becomes large since $\left\vert \nu _{\mathsf{N}%
}\right\vert >$\ $\left\vert \nu _{\mathsf{N+1}}\right\vert $\ and $\lim_{%
\mathsf{N}\rightarrow +\infty }\exp \left[ -\tau \left\vert \nu _{\mathsf{N}%
}\right\vert \right] =1$.

\bigskip

We can avoid the truncation method and yet obtain convergence results for
the solutions to (\ref{masterequation}) by modifying the basic argument, but
that is at the expense of having no error bounds in general unless we impose
additional conditions regarding the Fourier coefficients of the initial
data, as in Corollary 2 below. We begin with the crucial observation that (%
\ref{basicrelation}) still holds for an arbitrary initial condition $\mathsf{%
p}^{\ast }\in $ $l_{\mathbb{C},\mathsf{w}_{\beta ,\mu }}^{2}$ satisfying (%
\ref{probabilities}). More precisly we have:

\bigskip

\textbf{Lemma 1.} \textit{Let }$\mathsf{p}^{\ast }\in $\textit{\ }$l_{%
\mathbb{C},\mathsf{w}_{\beta ,\mu }}^{2}$\textit{\ satisfy the second
relation in (\ref{probabilities}). Then we have}%
\begin{equation*}
\left( \mathsf{p}^{\ast }\mathsf{,\hat{q}}_{\mathsf{1}}\right) _{2,\mathsf{w}%
_{\beta ,\mu }}\mathsf{\hat{q}}_{\mathsf{1}}=\mathsf{p}_{\beta \mathsf{,}\mu
}.
\end{equation*}

\bigskip

\textbf{Proof.} From (\ref{equiprobabilities}) and the definition of the
weights $w_{\beta ,\mu ,\mathsf{m}}$ we have 
\begin{equation*}
\left( \mathsf{p}_{\beta \mathsf{,}\mu },\mathsf{\hat{q}}_{\mathsf{k}%
}\right) _{2,\mathsf{w}_{\beta ,\mu }}=\Theta _{\beta ,\mu
}^{-1}\dsum\limits_{\mathsf{m=1}}^{\mathsf{+\infty }}\mathsf{\hat{q}}_{%
\mathsf{k,m}}=0
\end{equation*}%
for each $\mathsf{k\in }\left\{ 2,3,...\right\} $ by virtue of the
orthogonality of the eigenvectors of $A$, so that 
\begin{equation}
\dsum\limits_{\mathsf{m=1}}^{\mathsf{+\infty }}\mathsf{\hat{q}}_{\mathsf{k,m}%
}=0  \label{orthogonality}
\end{equation}%
for every such $\mathsf{k}$. Furthermore, for $\mathsf{p}^{\ast }\in $%
\textit{\ }$l_{\mathbb{C},\mathsf{w}_{\beta ,\mu }}^{2}$ we have the
norm-convergent series expansion%
\begin{equation*}
\mathsf{p}^{\ast }=\left( \mathsf{p}^{\ast }\mathsf{,\hat{q}}_{\mathsf{1}%
}\right) _{2,\mathsf{w}_{\beta ,\mu }}\mathsf{\hat{q}}_{\mathsf{1}%
}+\dsum\limits_{\mathsf{k=2}}^{\mathsf{+\infty }}\left( \mathsf{p}^{\ast }%
\mathsf{,\hat{q}}_{\mathsf{k}}\right) _{2,\mathsf{w}_{\beta ,\mu }}\mathsf{%
\hat{q}}_{\mathsf{k}}
\end{equation*}%
and therefore%
\begin{equation*}
\dsum\limits_{\mathsf{m=1}}^{\mathsf{+\infty }}\mathsf{p}_{\mathsf{m}}^{\ast
}=\left( \mathsf{p}^{\ast }\mathsf{,\hat{q}}_{\mathsf{1}}\right) _{2,\mathsf{%
w}_{\beta ,\mu }}\dsum\limits_{\mathsf{m=1}}^{\mathsf{+\infty }}\mathsf{\hat{%
q}}_{\mathsf{1,m}}=1
\end{equation*}%
as a consequence of (\ref{orthogonality}) and the second relation in (\ref%
{probabilities}). In this way (\ref{normalization}) holds again, so that we
may conclude as in the proof of Theorem 2. \ \ $\blacksquare $

\bigskip

Lemma 1 now allows us to get the following generalization of the preceding
theorem:

\bigskip

\textbf{Theorem 3. }\textit{Let }$A$\textit{\ be the operator defined by (%
\ref{operator}), and let }$\mathsf{p}^{\ast }\in l_{\mathbb{C},\mathsf{w}%
_{\beta ,\mu }}^{2}$ \textit{be any initial condition satisfying (\ref%
{probabilities}). Assuming moreover that (\ref{spectralcondition}) holds,
and that the ordering }$\nu _{\mathsf{k}}<\nu _{\mathsf{k+1}}$\textit{\ for
every} $\mathsf{k}\in \left\{ 2,3,...\right\} $ \textit{is still valid, we
have}%
\begin{equation*}
\lim_{\tau \rightarrow +\infty }\left\Vert \exp \left[ \tau A\right] \mathsf{%
p}^{\ast }-\mathsf{p}_{\beta \mathsf{,}\mu }\right\Vert _{2,\mathsf{w}%
_{\beta ,\mu }}\mathsf{=0.}
\end{equation*}

\bigskip

\textbf{Proof.} From the preceding lemma and its proof we may write%
\begin{equation*}
\mathsf{p}^{\ast }=\mathsf{p}_{\beta \mathsf{,}\mu }+\dsum\limits_{\mathsf{%
k=2}}^{\mathsf{+\infty }}\left( \mathsf{p}^{\ast }\mathsf{,\hat{q}}_{\mathsf{%
k}}\right) _{2,\mathsf{w}_{\beta ,\mu }}\mathsf{\hat{q}}_{\mathsf{k}}
\end{equation*}%
and therefore%
\begin{equation}
\left\Vert \exp \left[ \tau A\right] \mathsf{p}^{\ast }-\mathsf{p}_{\beta 
\mathsf{,}\mu }\right\Vert _{2,\mathsf{w}_{\beta ,\mu }}^{2}\mathsf{=}%
\dsum\limits_{\mathsf{k=2}}^{\mathsf{+\infty }}\left\vert \left( \mathsf{p}%
^{\ast }\mathsf{,\hat{q}}_{\mathsf{k}}\right) _{2,\mathsf{w}_{\beta ,\mu
}}\right\vert ^{2}\exp \left[ 2\tau \nu _{\mathsf{k}}\right] <+\infty
\label{quadraticformula}
\end{equation}%
for every $\tau \in \left[ 0,+\infty \right] $, without truncation. Now for
every fixed $\mathsf{k}\in \left\{ 2,3,...\right\} $ we have%
\begin{equation*}
\lim_{\tau \rightarrow +\infty }\left\vert \left( \mathsf{p}^{\ast }\mathsf{,%
\hat{q}}_{\mathsf{k}}\right) _{2,\mathsf{w}_{\beta ,\mu }}\right\vert
^{2}\exp \left[ 2\tau \nu _{\mathsf{k}}\right] =0
\end{equation*}%
since $\nu _{\mathsf{k}}<0$, and moreover%
\begin{equation*}
\left\vert \left( \mathsf{p}^{\ast }\mathsf{,\hat{q}}_{\mathsf{k}}\right)
_{2,\mathsf{w}_{\beta ,\mu }}\right\vert ^{2}\exp \left[ 2\tau \nu _{\mathsf{%
k}}\right] \leqslant \left\vert \left( \mathsf{p}^{\ast }\mathsf{,\hat{q}}_{%
\mathsf{k}}\right) _{2,\mathsf{w}_{\beta ,\mu }}\right\vert ^{2}
\end{equation*}%
for every $\mathsf{k}$ uniformly in $\tau \in \left[ 0,+\infty \right] $,
with%
\begin{equation*}
\dsum\limits_{\mathsf{k=2}}^{\mathsf{+\infty }}\left\vert \left( \mathsf{p}%
^{\ast }\mathsf{,\hat{q}}_{\mathsf{k}}\right) _{2,\mathsf{w}_{\beta ,\mu
}}\right\vert ^{2}\leqslant \left\Vert \mathsf{p}^{\ast }\right\Vert _{2,%
\mathsf{w}_{\beta ,\mu }}^{2}<+\infty \text{.}
\end{equation*}%
The result then follows from dominated convergence. \ \ $\blacksquare $

\bigskip

All the preceding results remain valid when $\mu =0$, which corresponds to
the description of a quantum system in thermodynamical equilibrium with a
heat bath at inverse temperature $\beta >0$, and to transition rates in (\ref%
{masterequation}) of the form%
\begin{equation}
r_{\mathsf{m,n}}=\exp \left[ -\frac{\beta }{2}\left( 3\lambda _{\mathsf{m}%
}+\lambda _{\mathsf{n}}\right) \right]  \label{ratester}
\end{equation}%
according to (\ref{ratesbis}). Furthermore, in this case the components (\ref%
{eigenvectorsbis}) of the eigenvectors of $A$ reduce to%
\begin{equation}
\hat{p}_{\mathsf{k,m}}=\frac{\exp \left[ -\frac{3\beta }{2}\lambda _{\mathsf{%
m}}\right] }{\nu _{\mathsf{k}}+b_{\mathsf{m}}}  \label{eigenvectorster}
\end{equation}%
where%
\begin{equation}
b_{\mathsf{m}}=Z_{\frac{3\beta }{2}}\exp \left[ -\frac{\beta }{2}\lambda _{%
\mathsf{m}}\right] .  \label{sequencebis}
\end{equation}%
In the preceding expression we have defined%
\begin{equation*}
Z_{\beta }:=\Theta _{\beta ,0}=\sum_{\mathsf{m=1}}^{+\infty }\exp \left[
-\beta \lambda _{\mathsf{m}}\right] <+\infty
\end{equation*}%
for every $\beta >0$, which stands for the usual partition function of the
canonical ensemble. Eigenvectors (\ref{eigenvectorster}) then constitute an
orthonormal basis of $l_{\mathbb{C},\mathsf{w}_{\beta }}^{2}$ where $\mathsf{%
w}_{\beta }:=\mathsf{w}_{\beta ,\mu =0}=\left( w_{\beta ,\mu =0,\mathsf{m}%
}\right) =\left( \exp \left[ \beta \lambda _{\mathsf{m}}\right] \right) $,
and moreover the grand canonical equilibrium distribution $\mathsf{p}_{\beta 
\mathsf{,}\mu }$ reduces to $\mathsf{p}_{\beta }:=\mathsf{p}_{\beta \mathsf{,%
}\mu =0}$ whose components are given by%
\begin{equation}
p_{\beta \mathsf{,m}}=Z_{\beta }^{-1}\exp \left[ -\beta \lambda _{\mathsf{m}}%
\right]  \label{gibbsdistribution}
\end{equation}%
for every $\mathsf{m}\in \mathbb{N}^{+}$. In the next result we state two
consequences of the above theorems:

\bigskip

\textbf{Corollary 1. }\textit{Let }$A$\textit{\ be the operator defined by (%
\ref{operator}), with the }$a_{\mathsf{m,n}}$ \textit{given by (\ref{matrix}%
) and (\ref{ratester})}. \textit{Then }$A:l_{\mathbb{C},\mathsf{w}_{\beta
}}^{2}\mapsto l_{\mathbb{C},\mathsf{w}_{\beta }}^{2}$ \textit{is a linear,
self-adjoint trace-class operator} \textit{whose trace is given by}%
\begin{equation*}
\func{Tr}A=Z_{2\beta }-Z_{\frac{\beta }{2}}Z_{\frac{3\beta }{2}}<0.
\end{equation*}%
\textit{Moreover, let us assume in addition that}%
\begin{equation}
\lambda _{\mathsf{m+1}}-\lambda _{\mathsf{m}}>0  \label{spectralconditionbis}
\end{equation}%
\textit{for every} $\mathsf{m}$\textsf{\ }$\in \mathbb{N}^{+}$, \textit{and
that the ordering }$\nu _{\mathsf{k}}<\nu _{\mathsf{k+1}}$\textit{\ of the
eigenvalues still holds for every} $\mathsf{k}\in \left\{ 2,3,...\right\} $. 
\textit{Then the following statements are valid:}

\textit{(a) For each }$\mathsf{N}\in \mathbb{N}^{+}$ \textit{with} $\mathsf{%
N\geqslant 2}$ \textit{and} $\mathsf{p}^{\ast }\in \vee _{\mathsf{k=1}}^{%
\mathsf{N}}E_{\nu _{\mathsf{k}}}\left( A\right) $ \textit{satisfying (\ref%
{probabilities}) we have the exponential decay estimate}%
\begin{equation*}
\left\Vert \exp \left[ \tau A\right] \mathsf{p}^{\ast }-\mathsf{p}_{\beta
}\right\Vert _{2,\mathsf{w}_{\beta }}\mathsf{\leqslant }\exp \left[ -\tau
\left\vert \nu _{\mathsf{N}}\right\vert \right] \left\Vert \mathsf{p}^{\ast
}\right\Vert _{2,\mathsf{w}_{\beta }}
\end{equation*}%
\textit{for every }$\tau \in \left[ 0,+\infty \right) $, \textit{where} $%
\mathsf{p}_{\beta }$ \textit{is given by (\ref{gibbsdistribution}).}

\textit{(b) Let }$\mathsf{p}^{\ast }\in l_{\mathbb{C},\mathsf{w}_{\beta
}}^{2}$ \textit{be any initial condition satisfying (\ref{probabilities}).
Then we have}%
\begin{equation*}
\lim_{\tau \rightarrow +\infty }\left\Vert \exp \left[ \tau A\right] \mathsf{%
p}^{\ast }-\mathsf{p}_{\beta }\right\Vert _{2,\mathsf{w}_{\beta }}\mathsf{=0.%
}
\end{equation*}

\bigskip

\textsc{Remark.} The operator $A$ of the preceding corollary may also be
realized as a non normal and non dissipative trace-class operator in the
usual unweighted Hilbert space $l_{\mathbb{C}}^{2}$ consisting of all square
summable complex sequences. This approach was implemented in \cite%
{boeglivuillermot}, with the goal of putting the analysis of $A$ into the
perspective of the spectral theory of linear non self-adjoint operators as
developed in \cite{gohbergkrein}. However, this was at the expense of having
to deal with a host of more complicated technical issues while imposing a
more restrictive condition on the spectral condition (\ref%
{spectralconditionbis}), namely,%
\begin{equation*}
\lambda _{\mathsf{m+1}}-\lambda _{\mathsf{m}}>c\exp \left[ -\theta \lambda _{%
\mathsf{m}}\right]
\end{equation*}%
for every $\mathsf{m}$\textsf{\ }$\in \mathbb{N}^{+}$, with both $c>0,\theta
>0$ independent of $\mathsf{m}$.

\bigskip

We complete this section by analyzing a concrete example which illustrates
the direct impact of the decay properties of the initial data in (\ref%
{masterequation}) on the speed of convergence of the corresponding
solutions. The example involves the quantum harmonic oscillator whose
spectrum we rescaled and shifted by irrelevant constants. We assume
throughout that $\mu =0$:

\bigskip

\textbf{Corollary 2. }\textit{Let us consider the initial-value problem (\ref%
{masterequation})-(\ref{probabilities}) where the transition rates are given
by (\ref{ratester})} \textit{and }$\lambda _{\mathsf{m}}=\mathsf{m}\in 
\mathbb{N}^{+}$. \textit{Moreover, let us assume that the ordering }$\nu _{%
\mathsf{k}}<\nu _{\mathsf{k+1}}$\textit{\ of the eigenvalues of the operator 
}$A$\textit{\ still holds for every} $\mathsf{k}\in \left\{ 2,3,...\right\} $%
. \textit{Then the following statements are valid:}

\textit{(a) If the Fourier coefficients of }$\mathsf{p}^{\ast }\in l_{%
\mathbb{C},\mathsf{w}_{\beta }}^{2}$\textit{\ along the orthonormal basis} $%
\left( \mathsf{\hat{q}}_{\mathsf{k}}\right) _{\mathsf{k}\in \mathbb{N}^{+}}$ 
\textit{of} $l_{\mathbb{C},\mathsf{w}_{\beta }}^{2}$ \textit{satisfy}%
\begin{equation}
\left\vert \left( \mathsf{p}^{\ast }\mathsf{,\hat{q}}_{\mathsf{k}}\right)
_{2,\mathsf{w}_{\beta }}\right\vert ^{2}\leqslant \kappa \exp \left[ -\delta 
\mathsf{k}\right]  \label{fourierdecay}
\end{equation}%
\textit{for every} $\mathsf{k}\in \left\{ 2,3,...\right\} $ \textit{and some}
$\kappa ,\delta >0$, \textit{then we have}%
\begin{equation}
\left\Vert \exp \left[ \tau A\right] \mathsf{p}^{\ast }-\mathsf{p}_{\beta
}\right\Vert _{2,\mathsf{w}_{\beta }}\leqslant c_{\beta ,\kappa ,\delta
}\tau ^{-\frac{\delta }{\beta }}  \label{polynomialdecay}
\end{equation}%
\textit{for all sufficiently large }$\tau $ \textit{and for some }$c_{\beta
,\kappa ,\delta }>0$\textit{\ depending solely on }$\beta ,\kappa $\textit{\
and }$\delta $\textit{.}

\textit{(b) If the Fourier coefficients of }$\mathsf{p}^{\ast }\in l_{%
\mathbb{C},\mathsf{w}_{\beta }}^{2}$\textit{\ along the orthonormal basis} $%
\left( \mathsf{\hat{q}}_{\mathsf{k}}\right) _{\mathsf{k}\in \mathbb{N}^{+}}$ 
\textit{of} $l_{\mathbb{C},\mathsf{w}_{\beta }}^{2}$ \textit{satisfy}%
\begin{equation}
\left\vert \left( \mathsf{p}^{\ast }\mathsf{,\hat{q}}_{\mathsf{k}}\right)
_{2,\mathsf{w}_{\beta }}\right\vert ^{2}\leqslant \kappa \mathsf{k}^{-\delta
}  \label{fourierdecaybis}
\end{equation}%
\textit{for every} $\mathsf{k}\in \left\{ 2,3,...\right\} $ \textit{and} 
\textit{some} $\kappa >0,$ $\delta >1$, \textit{then we have}%
\begin{equation}
\left\Vert \exp \left[ \tau A\right] \mathsf{p}^{\ast }-\mathsf{p}_{\beta
}\right\Vert _{2,\mathsf{w}_{\beta }}\leqslant c_{\beta ,\kappa ,\delta
}\left( \ln \tau \right) ^{-\frac{\delta -1}{2}}  \label{logarithmicdecay}
\end{equation}%
\textit{for all sufficiently large }$\tau $ \textit{and for some }$c_{\beta
,\kappa ,\delta }>0$\textit{\ depending solely on }$\beta ,\kappa $\textit{\
and }$\delta $\textit{.}

\bigskip

\textbf{Proof.} The starting point is the relation%
\begin{equation*}
\left\Vert \exp \left[ \tau A\right] \mathsf{p}^{\ast }-\mathsf{p}_{\beta
}\right\Vert _{2,\mathsf{w}_{\beta }}^{2}\mathsf{=}\dsum\limits_{\mathsf{k=2}%
}^{\mathsf{+\infty }}\left\vert \left( \mathsf{p}^{\ast }\mathsf{,\hat{q}}_{%
\mathsf{k}}\right) _{2,\mathsf{w}_{\beta }}\right\vert ^{2}\exp \left[ 2\tau
\nu _{\mathsf{k}}\right]
\end{equation*}%
which is (\ref{quadraticformula}) with $\mu =0$, where we assume that $\tau
>0$. Using (\ref{fourierdecay}) along with $\nu _{\mathsf{k}}<-b_{\mathsf{k}%
} $, the latter being a consequence of (\ref{localization}), we first obtain%
\begin{eqnarray}
&&\left\Vert \exp \left[ \tau A\right] \mathsf{p}^{\ast }-\mathsf{p}_{\beta
}\right\Vert _{2,\mathsf{w}_{\beta }}^{2}  \notag \\
&\leqslant &\kappa \dsum\limits_{\mathsf{k=2}}^{\mathsf{+\infty }}\exp \left[
-\delta \mathsf{k}-2\tau Z_{\frac{3\beta }{2}}\exp \left[ -\frac{\beta }{2}%
\mathsf{k}\right] \right]  \label{estimatester}
\end{eqnarray}%
by using (\ref{sequencebis}). In order to extract an explicit dependence in $%
\tau $ from the preceding expression let us now consider the function $%
f\left( .,\tau \right) :\left( 0,+\infty \right) \mapsto \mathbb{R}^{+}$
given by%
\begin{equation}
f(\mathsf{x,\tau }):=\exp \left[ -\delta \mathsf{x}-2\tau Z_{\frac{3\beta }{2%
}}\exp \left[ -\frac{\beta }{2}\mathsf{x}\right] \right] .  \label{function}
\end{equation}%
We remark that $f\left( .,\tau \right) $ possesses a unique critical point at%
\begin{equation}
\mathsf{x}_{c}(\tau )=\ln \left( \frac{c_{\beta }\tau }{\delta }\right) ^{%
\frac{2}{\beta }}  \label{criticalpoint}
\end{equation}%
where $c_{\beta }=\beta Z_{\frac{3\beta }{2}}$. Furthermore we choose $\tau $
sufficiently large so that the integer part of (\ref{criticalpoint})
satisfies $\left[ \mathsf{x}_{c}(\tau )\right] \geqslant 3$, with $f\left(
.,\tau \right) $ monotone increasing for $\mathsf{x}\in \left( 0,\mathsf{x}%
_{c}(\tau )\right) $ and monotone decreasing for $\mathsf{x}\in \left( 
\mathsf{x}_{c}(\tau ),+\infty \right) $. For the right-hand side of (\ref%
{estimatester}) we then obtain the estimate%
\begin{eqnarray}
&&\dsum\limits_{\mathsf{k=2}}^{\mathsf{+\infty }}f(\mathsf{k,\tau })  \notag
\\
&=&\dsum\limits_{\mathsf{k=2}}^{\left[ \mathsf{x}_{c}(\tau )\right] -1}f(%
\mathsf{k,\tau })+\dsum\limits_{\mathsf{k=}\left[ \mathsf{x}_{c}(\tau )%
\right] -1}^{\mathsf{+\infty }}f(\mathsf{k+1,\tau })  \label{estimatesquarto}
\\
&\leqslant &\int_{2}^{\left[ \mathsf{x}_{c}(\tau )\right] }d\mathsf{x}f(%
\mathsf{x,\tau })+f(\left[ \mathsf{x}_{c}(\tau )\right] \mathsf{,\tau })+f(%
\left[ \mathsf{x}_{c}(\tau )\right] +1\mathsf{,\tau })+\int_{\left[ \mathsf{x%
}_{c}(\tau )\right] +1}^{\mathsf{+\infty }}d\mathsf{x}f(\mathsf{x,\tau }) 
\notag \\
&\leqslant &\int_{\mathsf{2}}^{\mathsf{+\infty }}d\mathsf{x}f(\mathsf{x,\tau 
})+f(\left[ \mathsf{x}_{c}(\tau )\right] \mathsf{,\tau })+f(\left[ \mathsf{x}%
_{c}(\tau )\right] +1\mathsf{,\tau }).  \notag
\end{eqnarray}%
It is now easy to extract the desired dependence in $\tau $ for each term in
the preceding expression. For the integral this follows from the change of
variables $\mathsf{x\rightarrow y=\tau }\exp \left[ -\frac{\beta }{2}\mathsf{%
x}\right] ,$ which leads to the estimate 
\begin{eqnarray}
&&\int_{\mathsf{2}}^{\mathsf{+\infty }}d\mathsf{x}\exp \left[ -\delta 
\mathsf{x}-2\tau Z_{\frac{3\beta }{2}}\exp \left[ -\frac{\beta }{2}\mathsf{x}%
\right] \right]  \notag \\
&=&\frac{2}{\beta }\left( \int_{0}^{\tau \exp \left[ -\beta \right] }d%
\mathsf{yy}^{\frac{2\delta }{\beta }-1}\exp \left[ -2Z_{\frac{3\beta }{2}}%
\mathsf{y}\right] \right) \tau ^{-\frac{2\delta }{\beta }}  \notag \\
&\leqslant &\frac{2}{\beta }\left( \int_{0}^{+\infty }d\mathsf{yy}^{\frac{%
2\delta }{\beta }-1}\exp \left[ -2Z_{\frac{3\beta }{2}}\mathsf{y}\right]
\right) \tau ^{-\frac{2\delta }{\beta }}  \label{integrals} \\
&=&c_{\beta ,\delta }\Gamma \left( \frac{2\delta }{\beta }\right) \tau ^{-%
\frac{2\delta }{\beta }}  \notag
\end{eqnarray}%
for some $c_{\beta ,\delta }>0$ depending only on $\beta $ and $\delta $,
where $\Gamma $ stands for Euler's Gamma function.

As for the second and third terms on the right-hand side of (\ref%
{estimatesquarto}), we first note that the direct substitution of (\ref%
{criticalpoint}) into (\ref{function}) gives%
\begin{equation*}
f(\mathsf{x}_{c}(\tau )\mathsf{,\tau })=\hat{c}_{\beta ,\delta }\tau ^{-%
\frac{2\delta }{\beta }}
\end{equation*}%
where $\hat{c}_{\beta ,\delta }>0$, and therefore we get%
\begin{equation*}
f(\left[ \mathsf{x}_{c}(\tau )\right] \mathsf{,\tau })\leqslant \hat{c}%
_{\beta ,\delta }\tau ^{-\frac{2\delta }{\beta }}
\end{equation*}%
since $\left[ \mathsf{x}_{c}(\tau )\right] \leqslant \mathsf{x}_{c}(\tau )$
and since $f\left( .,\tau \right) $ is monotone increasing there. An
identical estimate holds for $f(\left[ \mathsf{x}_{c}(\tau )\right] +1%
\mathsf{,\tau })$ since $\mathsf{x}_{c}(\tau )<\left[ \mathsf{x}_{c}(\tau )%
\right] +1$ with $f\left( .,\tau \right) $ monotone decreasing there. The
substitution of all the gathered information into (\ref{estimatesquarto})
and the use of (\ref{estimatester}) then lead to (\ref{polynomialdecay}).

The proof of (\ref{logarithmicdecay}) follows a similar pattern but is a
little bit trickier. We start with%
\begin{eqnarray}
&&\left\Vert \exp \left[ \tau A\right] \mathsf{p}^{\ast }-\mathsf{p}_{\beta
}\right\Vert _{2,\mathsf{w}_{\mathsf{\beta }}}^{2}  \notag \\
&\leqslant &\kappa \dsum\limits_{\mathsf{k=2}}^{\mathsf{+\infty }}\mathsf{k}%
^{-\delta }\exp \left[ -2\tau Z_{\frac{3\beta }{2}}\exp \left[ -\frac{\beta 
}{2}\mathsf{k}\right] \right]  \label{estimatesquinto}
\end{eqnarray}%
and%
\begin{equation}
f(\mathsf{x,\tau }):=\mathsf{x}^{-\delta }\exp \left[ -2\tau Z_{\frac{3\beta 
}{2}}\exp \left[ -\frac{\beta }{2}\mathsf{x}\right] \right] .
\label{functionbis}
\end{equation}%
It is easily seen that the possible critical points of (\ref{functionbis})
are solutions to the equation%
\begin{equation}
\frac{\exp \left[ \frac{\beta }{2}\mathsf{x}\right] }{\mathsf{x}}=\frac{%
c_{\beta }\tau }{\delta }  \label{equation}
\end{equation}%
where $c_{\beta }$ is as in (\ref{criticalpoint}), and that the function on
the left-hand side of (\ref{equation}) is convex, possesses an absolute
minimum at $\mathsf{x}^{\ast }=\frac{2}{\beta }$ and is strictly increasing
for $\mathsf{x}\in \left( \mathsf{x}^{\ast },+\infty \right) $. Then for
every sufficiently large $\tau $ there exists a unique critical point $%
\mathsf{x}_{c}(\tau )\in \left( \mathsf{x}^{\ast },+\infty \right) $ of $%
f\left( .,\tau \right) $, this function being monotone increasing for $%
\mathsf{x}\in \left( \mathsf{x}^{\ast },\mathsf{x}_{c}(\tau )\right) $ and
monotone decreasing for $\mathsf{x}\in \left( \mathsf{x}_{c}(\tau ),+\infty
\right) $. Moreover, writing $\left[ \mathsf{x}^{\ast }\right] $ for the
integral part of $\mathsf{x}^{\ast }$, we may break up the right-hand side
of (\ref{estimatesquinto}) as%
\begin{eqnarray}
&&\dsum\limits_{\mathsf{k=2}}^{\mathsf{+\infty }}f(\mathsf{k,\tau })  \notag
\\
&=&\dsum\limits_{\mathsf{k=2}}^{\left[ \mathsf{x}^{\ast }\right] +2}f(%
\mathsf{k,\tau })+\dsum\limits_{\mathsf{k=}\left[ \mathsf{x}^{\ast }\right]
+3}^{\left[ \mathsf{x}_{c}(\tau )\right] -1}f(\mathsf{k,\tau }%
)+\dsum\limits_{\mathsf{k=}\left[ \mathsf{x}_{c}(\tau )\right] -1}^{\mathsf{%
+\infty }}f(\mathsf{k+1,\tau })  \label{estimatesdecimo} \\
&\leqslant &\dsum\limits_{\mathsf{k=2}}^{\left[ \mathsf{x}^{\ast }\right]
+2}f(\mathsf{k,\tau })+\int_{\left[ \mathsf{x}^{\ast }\right] +3}^{\mathsf{%
+\infty }}d\mathsf{x}f(\mathsf{x,\tau })+f(\left[ \mathsf{x}_{c}(\tau )%
\right] \mathsf{,\tau })+f(\left[ \mathsf{x}_{c}(\tau )+1\right] \mathsf{%
,\tau }).  \notag
\end{eqnarray}%
We now claim that the first term on the right-hand side of the preceding
inequality satisfies the exponential decay estimate%
\begin{equation}
\dsum\limits_{\mathsf{k=2}}^{\left[ \mathsf{x}^{\ast }\right] +2}f(\mathsf{%
k,\tau })\leqslant c_{\beta ,\delta }\exp \left[ -c_{\beta }\tau \right]
\label{estimatessexto}
\end{equation}%
for some $c_{\beta ,\delta },c_{\beta }>0$. Indeed we have%
\begin{eqnarray*}
&&\dsum\limits_{\mathsf{k=2}}^{\left[ \mathsf{x}^{\ast }\right] +2}\mathsf{k}%
^{-\delta }\exp \left[ -2\tau Z_{\frac{3\beta }{2}}\exp \left[ -\frac{\beta 
}{2}\mathsf{k}\right] \right] \\
&\leqslant &\mathsf{2}^{-\delta }\left( \left[ \mathsf{x}^{\ast }\right]
+1\right) \exp \left[ -2\tau Z_{\frac{3\beta }{2}}\exp \left[ -\frac{\beta }{%
2}\left( \left[ \mathsf{x}^{\ast }\right] +2\right) \right] \right]
\end{eqnarray*}%
since $2\leqslant \mathsf{k}\leqslant \left[ \mathsf{x}^{\ast }\right] +2$,
which is (\ref{estimatessexto}) with an obvious choice for $c_{\beta ,\delta
}$ and $c_{\beta }$ as $\left[ \mathsf{x}^{\ast }\right] $ depends only on $%
\beta $.

As for the integral we have%
\begin{eqnarray*}
&&\int_{\left[ \mathsf{x}^{\ast }\right] +3}^{\mathsf{+\infty }}d\mathsf{xx}%
^{-\delta }\exp \left[ -2\tau Z_{\frac{3\beta }{2}}\exp \left[ -\frac{\beta 
}{2}\mathsf{x}\right] \right] \\
&=&c_{\beta ,\delta }\int_{0}^{\tau \exp \left[ -\frac{\beta }{2}\left( %
\left[ \mathsf{x}^{\ast }\right] +3\right) \right] }\frac{d\mathsf{y}}{%
\mathsf{y}}\left( \ln \frac{\tau }{\mathsf{y}}\right) ^{-\delta }\exp \left[
-2Z_{\frac{3\beta }{2}}\mathsf{y}\right]
\end{eqnarray*}%
following the same change of variables as in (\ref{integrals}), for some $%
c_{\beta ,\delta }>0$. Therefore, integrating by parts and using the fact
that $\delta >1$ to control the completely integrated term we obtain,
changing the value of $c_{\beta ,\delta }$ if necessary, 
\begin{eqnarray}
&&\int_{\left[ \mathsf{x}^{\ast }\right] +3}^{\mathsf{+\infty }}d\mathsf{x}f(%
\mathsf{x,\tau })  \label{estimatesseptimo} \\
&=&c_{\beta ,\delta }\exp \left[ -c_{\beta }\tau \right] +\hat{c}_{\beta
,\delta }\int_{0}^{\tau \exp \left[ -\frac{\beta }{2}\left( \left[ \mathsf{x}%
^{\ast }\right] +3\right) \right] }d\mathsf{y}\left( \ln \frac{\tau }{%
\mathsf{y}}\right) ^{1-\delta }\exp \left[ -2Z_{\frac{3\beta }{2}}\mathsf{y}%
\right]  \notag
\end{eqnarray}%
for some $c_{\beta },\hat{c}_{\beta ,\delta }>0$, thereby exhibiting the
exponential decay of the first term on the right-hand side. In order to
extract the dependence in $\tau $ of the second term we start with 
\begin{eqnarray*}
&&\int_{0}^{\tau \exp \left[ -\frac{\beta }{2}\left( \left[ \mathsf{x}^{\ast
}\right] +3\right) \right] }d\mathsf{y}\left( \ln \frac{\tau }{\mathsf{y}}%
\right) ^{1-\delta }\exp \left[ -2Z_{\frac{3\beta }{2}}\mathsf{y}\right] \\
&=&\int_{0}^{\sqrt{\tau }}d\mathsf{y}\left( \ln \frac{\tau }{\mathsf{y}}%
\right) ^{1-\delta }\exp \left[ -2Z_{\frac{3\beta }{2}}\mathsf{y}\right] \\
&&+\int_{\sqrt{\tau }}^{\tau \exp \left[ -\frac{\beta }{2}\left( \left[ 
\mathsf{x}^{\ast }\right] +3\right) \right] }d\mathsf{y}\left( \ln \frac{%
\tau }{\mathsf{y}}\right) ^{1-\delta }\exp \left[ -2Z_{\frac{3\beta }{2}}%
\mathsf{y}\right] ,
\end{eqnarray*}%
which leads to the estimate%
\begin{eqnarray}
&&\int_{0}^{\sqrt{\tau }}d\mathsf{y}\left( \ln \frac{\tau }{\mathsf{y}}%
\right) ^{1-\delta }\exp \left[ -2Z_{\frac{3\beta }{2}}\mathsf{y}\right] 
\notag \\
&\leqslant &c_{\delta }\left( \int_{0}^{+\infty }d\mathsf{y}\exp \left[ -2Z_{%
\frac{3\beta }{2}}\mathsf{y}\right] \right) \left( \ln \tau \right)
^{1-\delta }=c_{\beta ,\delta }\left( \ln \tau \right) ^{1-\delta }
\label{estimatesoitavo}
\end{eqnarray}%
for the first term on the right-hand side where $c_{\delta },c_{\beta
,\delta }>0$. As for the second term we get%
\begin{eqnarray}
&&\int_{\sqrt{\tau }}^{\tau \exp \left[ -\frac{\beta }{2}\left( \left[ 
\mathsf{x}^{\ast }\right] +3\right) \right] }d\mathsf{y}\left( \ln \frac{%
\tau }{\mathsf{y}}\right) ^{1-\delta }\exp \left[ -2Z_{\frac{3\beta }{2}}%
\mathsf{y}\right]  \notag \\
&\leqslant &\left( \int_{\sqrt{\tau }}^{\tau \exp \left[ -\frac{\beta }{2}%
\left( \left[ \mathsf{x}^{\ast }\right] +3\right) \right] }d\mathsf{y}\left(
\ln \frac{\tau }{\mathsf{y}}\right) ^{1-\delta }\right) \exp \left[ -2Z_{%
\frac{3\beta }{2}}\sqrt{\tau }\right]  \notag \\
&=&\left( \int_{\frac{1}{\sqrt{\tau }}}^{\exp \left[ -\frac{\beta }{2}\left( %
\left[ \mathsf{x}^{\ast }\right] +3\right) \right] }d\mathsf{y}\left( \ln 
\frac{1}{\mathsf{y}}\right) ^{1-\delta }\right) \tau \exp \left[ -2Z_{\frac{%
3\beta }{2}}\sqrt{\tau }\right]  \notag \\
&\leqslant &\left( \int_{0}^{\exp \left[ -\frac{\beta }{2}\left( \left[ 
\mathsf{x}^{\ast }\right] +3\right) \right] }d\mathsf{y}\left( \ln \frac{1}{%
\mathsf{y}}\right) ^{1-\delta }\right) \tau \exp \left[ -2Z_{\frac{3\beta }{2%
}}\sqrt{\tau }\right]  \notag \\
&=&\hat{c}_{\beta ,\delta }\tau \exp \left[ -2Z_{\frac{3\beta }{2}}\sqrt{%
\tau }\right]  \label{estimatesnono}
\end{eqnarray}%
for some $\hat{c}_{\beta ,\delta }>0$, the last improper integral being
convergent. The combination of (\ref{estimatesseptimo})-(\ref{estimatesnono}%
) thus leads to%
\begin{equation}
\int_{\left[ \mathsf{x}^{\ast }\right] +3}^{\mathsf{+\infty }}d\mathsf{x}f(%
\mathsf{x,\tau })\leqslant c_{\beta ,\delta }\left( \ln \tau \right)
^{1-\delta }  \label{estimatesonzimo}
\end{equation}%
for some appropriate $c_{\beta ,\delta }>0$.

It remains to estimate the last two terms on the right-hand side of
inequality (\ref{estimatesdecimo}). We begin by observing that (\ref%
{functionbis}) implies%
\begin{equation}
f(\mathsf{x}_{c}(\tau )\mathsf{,\tau })\leqslant \mathsf{x}_{c}^{-\delta
}(\tau )  \label{equationbis}
\end{equation}%
for every $\tau \in \left( 0,+\infty \right) $, while (\ref{equation}) and
the fact that $\mathsf{x}_{c}(\tau )>\mathsf{x}^{\ast }$ for $\tau $
sufficiently large lead to%
\begin{equation*}
\exp \left[ \frac{\beta }{2}\mathsf{x}_{c}(\tau )\right] =\mathsf{x}%
_{c}(\tau )\frac{c_{\beta }\tau }{\delta }\geqslant \mathsf{x}^{\ast }\frac{%
c_{\beta }\tau }{\delta }.
\end{equation*}%
Since $\mathsf{x}^{\ast }=\frac{2}{\beta }$, we may therefore change the
value of $c_{\beta }$ if necessary and thus obtain the lower bounds%
\begin{equation}
\mathsf{x}_{c}(\tau )\geqslant \frac{2}{\beta }\ln \frac{c_{\beta }\tau }{%
\delta }\geqslant \frac{1}{\beta }\ln \tau  \label{inequality}
\end{equation}%
for the critical point, where the second inequality follows from the fact
that we may take $\frac{c_{\beta }\tau }{\delta }>\sqrt{\tau }$ for $\tau $
sufficiently large since $\frac{c_{\beta }}{\delta }>0$. From (\ref%
{equationbis}) and (\ref{inequality}) we then get 
\begin{equation*}
f(\mathsf{x}_{c}(\tau )\mathsf{,\tau })\leqslant c_{\beta ,\delta }\left(
\ln \tau \right) ^{-\delta }
\end{equation*}%
for some suitably chosen $c_{\beta ,\delta }>0$, so that arguing as in the
proof of Statement (a) we end up with 
\begin{equation*}
f(\left[ \mathsf{x}_{c}(\tau )\right] \mathsf{,\tau })\leqslant c_{\beta
,\delta }\left( \ln \tau \right) ^{-\delta }
\end{equation*}%
and with an identical bound for $f(\left[ \mathsf{x}_{c}(\tau )\right] +1%
\mathsf{,\tau })$. The substitution of this information along with (\ref%
{estimatessexto}) and (\ref{estimatesonzimo}) into (\ref{estimatesdecimo})
then leads (\ref{logarithmicdecay}). \ \ $\blacksquare $

\bigskip

\textsc{Remark.} Throughout this article we carried out our computations
with transition rates given by (\ref{rates}) and (\ref{suitableconstants})
mainly for the sake of clarity and simplicity. However, there are plenty of
other choices for them that lead to similar results, as long as they satisfy
the detailed balance conditions (\ref{detailedbalancebis}). Furthermore, as
an illustration of our considerations we showed in Corollary 2 that even if
the quantum harmonic oscillator is initially steered away from thermodynamic
equilibrium due to its interaction with a heat bath at inverse temperature $%
\beta >0$, it will eventually return there at a rate which strongly depends
on the decay properties of the initial conditions (\ref{probabilities}), a
result that is complementary to those in Section 3 of \cite{daviesbis}.

\bigskip

\textbf{Acknowledgements.} The second author would like to thank the Funda%
\c{c}\~{a}o para a Ci\^{e}ncia e a Tecnologia of the Portuguese Government
(FCT) for its financial support under grant PDTC/MAT-STA/0975/2014.

\end{document}